\documentclass{pasj00}
\draft

\begin{document}
\SetRunningHead{Shimizu et al.}{Flare associated high-speed photospheric 
flow in a $\delta$-type sunspot}

\title{High speed photospheric material flow observed at the polarity inversion line of 
a $\delta$-type sunspot producing an X5.4 flare on 7 March 2012}

\author{Toshifumi \textsc{Shimizu} %
  }
\affil{Institute of Space and Astronautical Science,
Japan Aerospace Exploration Agency, 
3-1-1 Yoshinodai, Chuo-ku, Sagamihara, Kanagawa 252-5210, Japan.}
\email{shimizu@solar.isas.jaxa.jp}

\author{Bruce W. \textsc{Lites}}
\affil{High Altitude Observatory, National Center for
Atmospheric Research, P.O.Box 3000, Boulder, CO 80307, 
USA.\thanks{The National Center for Atmospheric Research is
sponsored by the National Science Foundation.}}
\email{lites@ucar.edu}
\and
\author{Yumi {\sc Bamba}}
\affil{Solar-Terrestrial Environment Laboratory, Nagoya University, Furo-cho, Chikusa-ku, Nagoya,
Aichi 464-8601, Japan.
\thanks{Also at Institute of Space and Astronautical Science,
Japan Aerospace Exploration Agency, 
3-1-1 Yoshinodai, Chuo-ku, Sagamihara, Kanagawa 252-5210, Japan.}}
\email{y-bamba@stelab.nagoya-u.ac.jp}

%


\maketitle

\begin{abstract}

Solar flares abruptly release the free energy stored as 
a non-potential magnetic field in the corona and may be
accompanied by eruptions of the coronal plasma. 
Formation of a non-potential magnetic field and the mechanisms
for triggering the onset of flares are still poorly understood.
In particular, 
photospheric dynamics
observed near those polarity inversion lines 
that are sites of major flare production have not been well observed
with high spatial resolution spectro-polarimetry. 
This paper reports on a remarkable high-speed material flow
observed along the polarity inversion line located between 
flare ribbons at the main energy release side of an
X5.4 flare on 7 March 2012. Observations were carried out
by the spectro-polarimeter of the Solar Optical Telescope onboard
{\em Hinode}. The high-speed material flow was observed in
the horizontally-oriented magnetic field formed nearly
parallel to the polarity inversion line.  This flow persisted from 
at least 6 hours before the onset of the flare, and 
continued for at least several hours after the onset of
the flare. Observations suggest that the observed material
flow represents neither the emergence
nor convergence of the magnetic flux. Rather, it may be 
considered to be material flow working both to increase
the magnetic shear along the polarity inversion line
and to develop magnetic structures favorable 
for the onset of the eruptive flare. 

\end{abstract}

\section{Introduction}

Solar flares and their associated coronal mass ejections, the most energetic bursts 
in our solar system,
abruptly release 
the free energy stored as a non-potential magnetic field in the corona and 
may be accompanied by eruptions of the coronal plasma, giving rise to
abrupt, widespread, and sometimes severe 
alterations of the solar-terrestrial environment. 
The formation in the corona of an S-shaped structure called a sigmoid \citep{rus96, can99}
is one indicator of a non-potential magnetic field that may lead
to eruptive major flares. 
Sigmoids are formed in the corona, associated with temporal evolution 
of magnetic flux at the photosphere including footpoint
motions of magnetic flux and the emergence of twisted flux from 
below the photosphere.
Most theoretical studies assume one of two typical motions at the photosphere
for developing a non-potential field in the corona: shear motions 
along polarity inversion line or converging motions toward 
the polarity inversion line. 
With such motions, a helically-twisted flux 
rope
is 
created in the corona by reconnection along the polarity inversion line.
This flux rope may be 
inferred to be the magnetic structure that supports dense filament plasma
in the active region
\citep{van89, ama00}. 
An alternate model is the emergence of twisted flux from below 
the photosphere, forming a non-potential structure in the corona \citep{mag01, arc12}.
In this scenario, observations of the footpoint motions of the magnetic
field at  the photosphere are necessary to identify motions and dynamical behavior 
that are responsible for building up a non-potential field in the corona.
At the same time, footpoint motions and dynamical behavior of the magnetic field 
play key roles in triggering the onset of major flares. In most numerical 
simulations, applying either a shear or a converging motion to the 
photospheric magnetic field, or injecting
emerging flux at the photosphere, 
act as
a trigger for the onset of mass
ejections and flares.  High resolution and accurate observations 
for identifying these motions and dynamical behavior are still limited. 

Most of major flares (X and $>$M5 flares) are produced in $\delta$-type sunspots \citep{sam00}. 
The $\delta$-type is one of Hale's sunspot magnetic classes.  It
has a penumbra enclosing umbrae of both positive and negative polarity. 
High magnetic shear is sometimes observed at the polarity inversion line
that, by definition, is present in $\delta$-type sunspots. 
The polarity inversion line is the locus where 
radial field reverses direction.
The magnetic field near the polarity inversion line may show complicated
configurations with steady and/or dynamic material flows.
\citet{mar94} reported extremely large downward-directed Doppler velocities 
in the vicinity of the polarity inversion line of a $\delta$-type sunspot.
\citet{lit02} reported Doppler shifts implying high-velocity, steady flows 
converging upon the line separating opposite polarity magnetic fields.  These flows
were interpreted within the context of the Evershed flow observed in all sunspots. 
\citet{tak12} detected continuous prominent downflows of $1.5 -1.7$ km s$^{-1}$
for several hours at the polarity inversion line.
Moreover, there are some reports on remarkable long-lived flows 
associated with the occurrence of major flares, although reports so far
are rare. 
\citet{meu03} observed long-lived,
highly inclined supersonic photospheric downflows and possible 
shear flows in a flaring active region.
\citet{yan04} and \citet{den06} also observed
long-lived (at least 5 hr) strong horizontal and vertical shear flows
(both in the order of 1 km s$^{-1}$) along the polarity inversion line until an X-class flare 
occurred. Both horizontal and vertical shear flows enhanced dramatically
after that flare, suggesting that photospheric shear flows and local
magnetic shear near the polarity inversion can increase after the flare
as the result of shear release in the overlying large-scale magnetic system.

These Doppler velocities are observed in  regions of complicated magnetic field.
Therefore, accurate spectro-polarimetry that provides quantitative measures of the
magnetic field vector is necessary to extract information 
about flow field in the small-scale magnetic field structure. 
There is a high incidence of major flares in the vicinity of the
polarity inversion lines of $\delta$-type sunspots, therefore 
 the dynamical behavior giving rise to Doppler shifts revealed by spectro-polarimetry
may provide hints toward understanding the build-up and
trigger process of flares. In this paper, we investigate
Doppler velocity and magnetic fields around 
the polarity inversion line in a sequence of three spectro-polarimetric
maps acquired in
an 8 hour period during which a major flare took place. 
Section \ref{sec: obs} describes observations and  data
analysis. After identifying the main energy release site of the flare
in Section \ref{sec: flare}, we present Doppler velocity and magnetic field maps at
the energy release site in Section \ref{sec: results}. 
The energy release site has  a polarity inversion line between the chromospheric flare ribbons,
where a remarkable high speed photospheric flow is found 
along the polarity inversion line. 
We discuss the results in section \ref{sec: discussions}  and finally conclude 
in section \ref{sec: summary}.

\section{Observations and data analysis}
\label{sec: obs}

\begin{figure}
 \begin{center}
  \includegraphics[width=12cm]{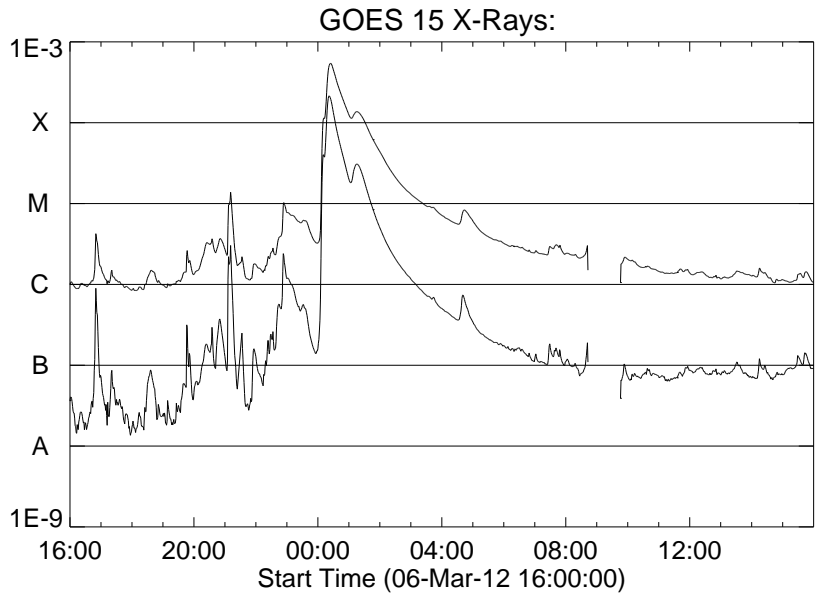} 
 \end{center}
\caption{GOES-15 soft X-ray flux plot on 6-7 March 2012. The upper and lower lines
indicate the full-sun soft X-ray flux through 1-8\AA\ and 0.5-4\AA\ , respectively.
The vertical axis gives the flare classes (X, M, C, B, and A) with the flux values
at the both ends of the figure in Watt m$^{-2}$.}
\label{fig: goes}
\end{figure}

An X5.4 flare was produced on 7 March 2012 from NOAA Active
Region 11429, which was a complex of sunspots including $\delta$-type
sunspots. Figure~\ref{fig: goes} shows the time profiles of soft X-ray 
flux from {\em GOES}. The soft X-ray flux of this X5.4 flare began to 
increase at 00:00 UT, and peaked at
00:24 UT on 7 March 2012. The location of the flare
was reported to be N17 E27 on the solar disk. The flare was followed by an X1.3 flare
whose peak at 1:14 UT was located at N22 E12. 
The coronal mass ejection launched at the same time as the X5.4 flare 
propagated through interplanetary space and
caused a large geomagnetic storm on 9 March \citep{tsu14}.

The Solar Optical Telescope (SOT) \citep{tsu08, sue08, shi08, ich08}
onboard {\em Hinode} \citep{kos07} 
observed
this active region and acquired Stokes spectral profiles 
with 
the
spectro-polarimeter (SP) for monitoring the temporal evolution 
of photospheric magnetic field after 10 UT, 6 March 2012.  
The SP \citep{lit13} records the full-polarization states 
of line profiles of two magnetically sensitive Fe I lines at 630.15 
and 630.25 nm. The fast-mapping mode was used, which
covers the entire sunspot with 0.32" effective pixel size. 
The spectral sampling is 21.549 m\AA\ pixel$^{-1}$. 

In this study, we analyzed three SP maps in detail.
Two of the maps were recorded at 17:30-18:02 UT and 22:10-22:43 UT,
6 March 2012, i.e., about 6 hours and 1.5 hours before the onset of
the X5.4 flare, respectively. The last map was recorded at 2:21-2:54 UT,  7 March
2012, i.e., about 2 hours after the onset of the X5.4 flare. 
We utilized the SOT/SP level2 database, which 
are outputs from inversions using the MERLIN inversion code
developed under the Community Spectro-polarimetric Analysis Center (CSAC) 
initiative (http://www.csac.hao.ucar.edu/) at HAO/NCAR. 
The inversion code performs a least-squares
fitting of the Stokes profiles using the Milne-Eddington atmospheric 
approximation that allows for a linear variation of the source function 
along the line-of-sight, but holds the magnetic field vector, line
strength, Doppler shift, line broadening, magnetic fill fraction constant 
along the line-of-sight. 

The 180\degree~ambiguity in the azimuth angle of 
magnetic field was resolved by using the AZAM 
utility \citep{lit95}, where the azimuth was selected to minimize 
spatial discontinuities in the field orientation. Since magnetic
field configuration near the polarity inversion line is complex
and there is some subjectivity to the ``disambiguation'', two of
the authors obtained the 180\degree~ambiguity resolution independently
and verified consistency of the two results. 
The magnetic field presented in this paper is in the local solar frame,
i.e., the coordinate transformed as if viewed from directly above. 
The position on the solar disk was used to transform the vector magnetic field 
from the observing (line-of-sight) frame coordinates to local solar frame coordinates. 

Line-of-sight magnetograms from the Helioseismic and Magnetic Imager
(HMI; \cite{sche12, scho12}) on board Solar Dynamics Observatory
(SDO; \cite{pes12}) were also examined for monitoring the temporal 
evolution of photospheric magnetic flux at the polarity inversion line.
HMI observes the full solar disk at 
617.3 nm
with a spatial resolution of 1"
and produces line-of-sight magnetograms in a cadence of 45 sec. 
We utilized the HMI level2 database available from
the Joint Science Operations Center at Stanford. 

\begin{figure}
 \begin{center}
  \includegraphics[width=12cm]{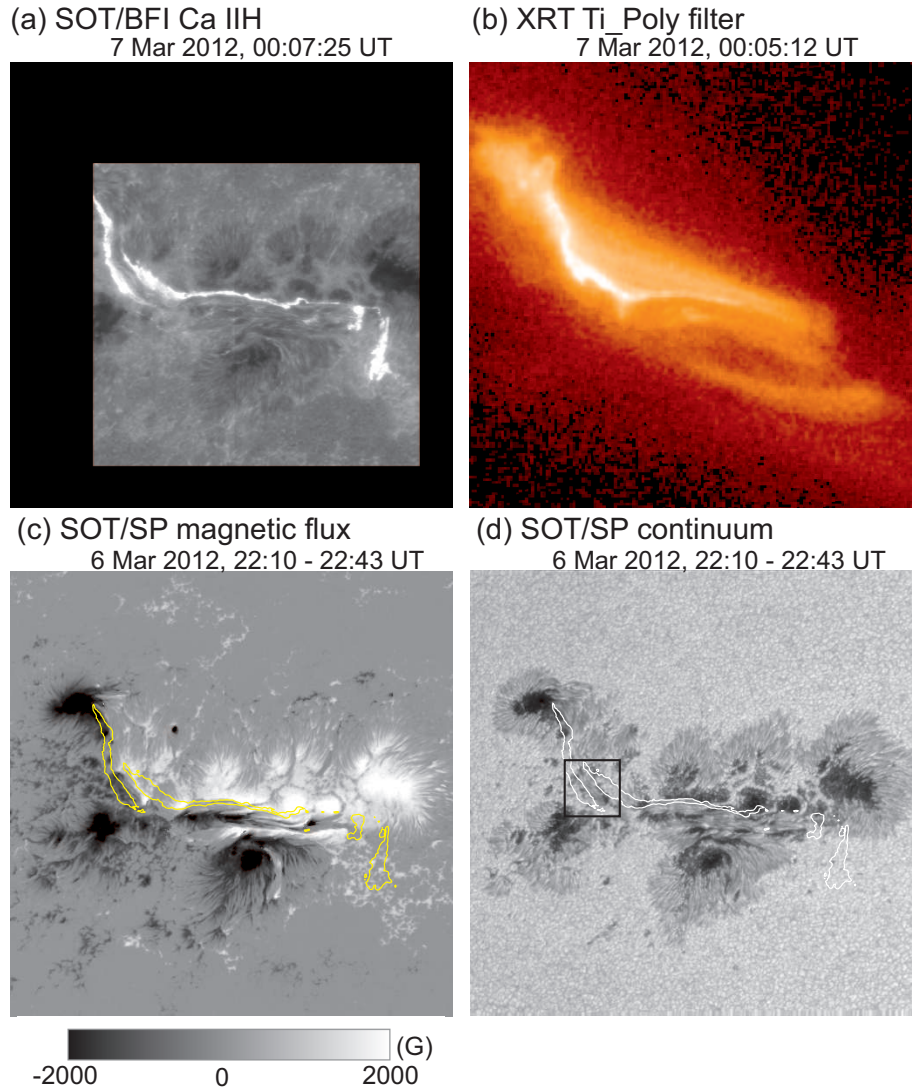} 
 \end{center}
\caption{Active region 11429 in (a) Ca II H, (b) soft X-rays, (c) vertical component
of magnetic flux density Bz at the photosphere, 
and (d) continuum. The Ca II H image was taken with SOT BFI 
just after the onset of the X5.4 flare. The contours in (c) and (d) represent
the position of the flare ribbons seen in the Ca II H image. 
The Bz and continuum images are produced from a fast mapping observation of
the SOT spectro-polarimeter at 22:10-22:43 UT on 6 March 2012. The soft x-ray image
was from {\em Hinode} XRT. The field of view is $160 \times 160$ arc sec.
North is up and west is to the right.
The square (about $20 \times 20$ arc sec) gives the field of view in the subsequent
figures. }
\label{fig: ar}
\end{figure}

From simultaneous time series of chromospheric and X-ray imaging obtained by
{\em Hinode}, Figure~\ref{fig: ar} shows a Ca {\sc II H} image recorded 
with  the SOT broadband imager (BFI)
at 00:07:25 UT on 7 March 2012 and a soft X-ray image (Ti Poly filter) from
the {\em Hinode} X-ray Telescope (XRT) \citep{gol07, kan08}.   These image sequences
were used to identify 
the footpoints of flaring arcade loops and the main energy release site
of the flare.

\section{Energy release site of X5.4 flare}
\label{sec: flare}

Figure~\ref{fig: ar}(a) shows the location of flare ribbons seen in the chromospheric
Ca II H image just after the time of the flare onset. Two bright ribbons appeared 
with a short ribbon at the left and a longer ribbon at the right. The short ribbon
is located in the negative polarity region and the long ribbon is in the positive polarity side.
The spatial arrangement of the flare ribbons indicates that sheared magnetic field in
the corona is involved in the main energy release of the X5.4 flare. 
The series of soft X-ray images also show 
the presence of a sigmoid
and a bright flaring X-ray source appeared after the onset of the flare.
The initial phase of the bright flaring X-ray source is seen in Figure~\ref{fig: ar}(b).
The bright source was confined, and existed along the polarity inversion line. 
The brightest portion of the bright X-ray source is located in the square (about $20 \times 20$ arc sec) 
given in Figure~\ref{fig: ar}(d).  In the square, the separation between the flare
ribbons in Ca II H is about 10 arc sec, and the polarity inversion line is located
almost parallel to and between the flare ribbons.
In subsequent sections, we will investigate the photospheric 
magnetic field and line-of-sight velocity field in the square for understanding
the formation and trigger of the X5.4 flare. 

It is noted that the energy release site of the following X1.3 flare differs slightly
from that of the X5.4 flare.  
The bright source of the X5.4 flare gradually evolved to an arcade-like
coronal structure at the left half of the active region. 
At the time of the X1.3 flare, a low-lying sheared coronal structure 
at the right half of active region suddenly brightened and 
formed an X-ray arcade. 

\section{High speed photospheric material flow}
\label{sec: results}
 
\begin{figure}
 \begin{center}
  \includegraphics[width=12cm]{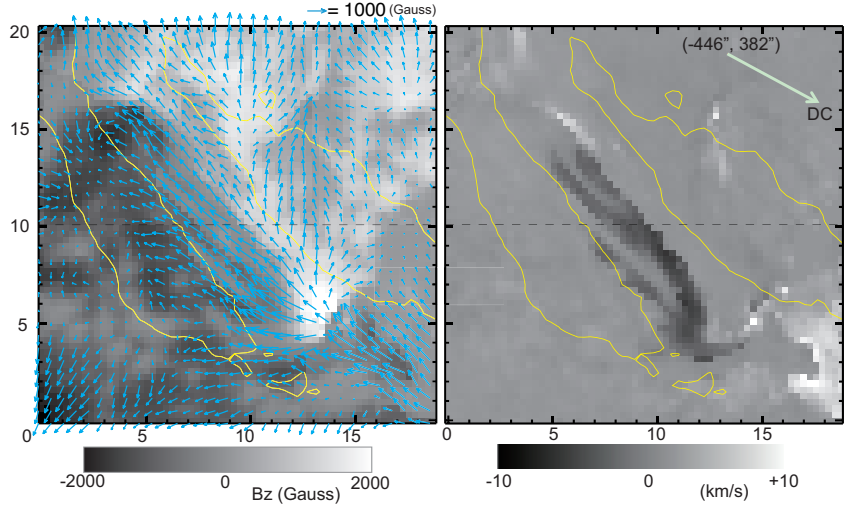} 
 \end{center}
\caption{Spatial distribution of the vector magnetic field (left) and Doppler
signals (right) in 
the square given in Figure~\ref{fig: ar}(d). 
The size of the square is about $20 \times 20$ arc sec.
The parameters were derived from the SP map acquired at
 22:10 - 22:43 UT, 6 March 2012. On the vector magnetic field map,
the aqua-colored arrows show the horizontal component ($Bx$, $By$) of magnetic fields, and
the background gray scale indicates the strength of the vertical component of the field ($Bz$). 
Redshift is positive in the Doppler signal map.
The dashed line is for Figure~\ref{fig: slice}, which shows
the spatial profiles of the Doppler velocity and magnetic
field quantities.
Yellow contours  show the approximate position of the flare ribbons as
determined from the Ca II H image at 00:07:25 UT, 7 March 2012. 
The region is located at (-446, 382) arc sec on the solar disk and 
the direction to the disk center is indicated. 
}
\label{fig: doppler}
\end{figure}

Figure~\ref{fig: doppler} shows the photospheric magnetic field and 
line-of-sight velocity field at the main energy release site, i.e., 
in the square given in Figure~\ref{fig: ar}. 
In the left panel, arrows give the horizontal component of the vector magnetic field ($Bx$, $By$) 
in the local solar frame. 
The background image is the vertical component of the vector magnetic field
($Bz$) and the positive and negative polarity flux is represented by white and black, 
respectively. The data show that the horizontally-oriented field is dominant 
in the region between the flare ribbons at the main energy release site,
and that it is 
nearly parallel to
the polarity inversion line located between the flare ribbons.  
At the both ends of the horizontal field, compact positive- and negative-polarity 
islands exist. Since the positive island at the SW 
[(13, 5) arc sec in Figure~\ref{fig: doppler}] is located 
on the positive side of the polarity inversion line and the negative island at NE 
[(3, 15) arc sec in Figure~\ref{fig: doppler}] is 
on the negative side, the horizontal field crossing
the polarity inversion line may be considered as having high magnitude of shear. 

The right panel of Figure~\ref{fig: doppler} 
shows
the spatial distribution of
Doppler signals at the photosphere for the main energy release site. 
Negative Doppler signals are recognized in majority of the horizontal field.
The speed is a few km s$^{-1}$ with supersonic speed over 5 km s$^{-1}$ at some pixels. 
Negative means blue-shifted, i.e., Doppler shift toward the observer. 
Since the region is located at (-446, 382) arc sec on the NE quadrant
of the solar disk, the direction normal to the solar surface is 
tilted about 30\degree~from the line-of-sight direction. 
Therefore, the observed negative Doppler signals imply 
one-directional material flow from the negative-polarity island 
to the positive-polarity island under the assumption that the Doppler velocity vector is 
aligned with the magnetic field vector. 

\begin{figure}
 \begin{center}
  \includegraphics[width=12cm]{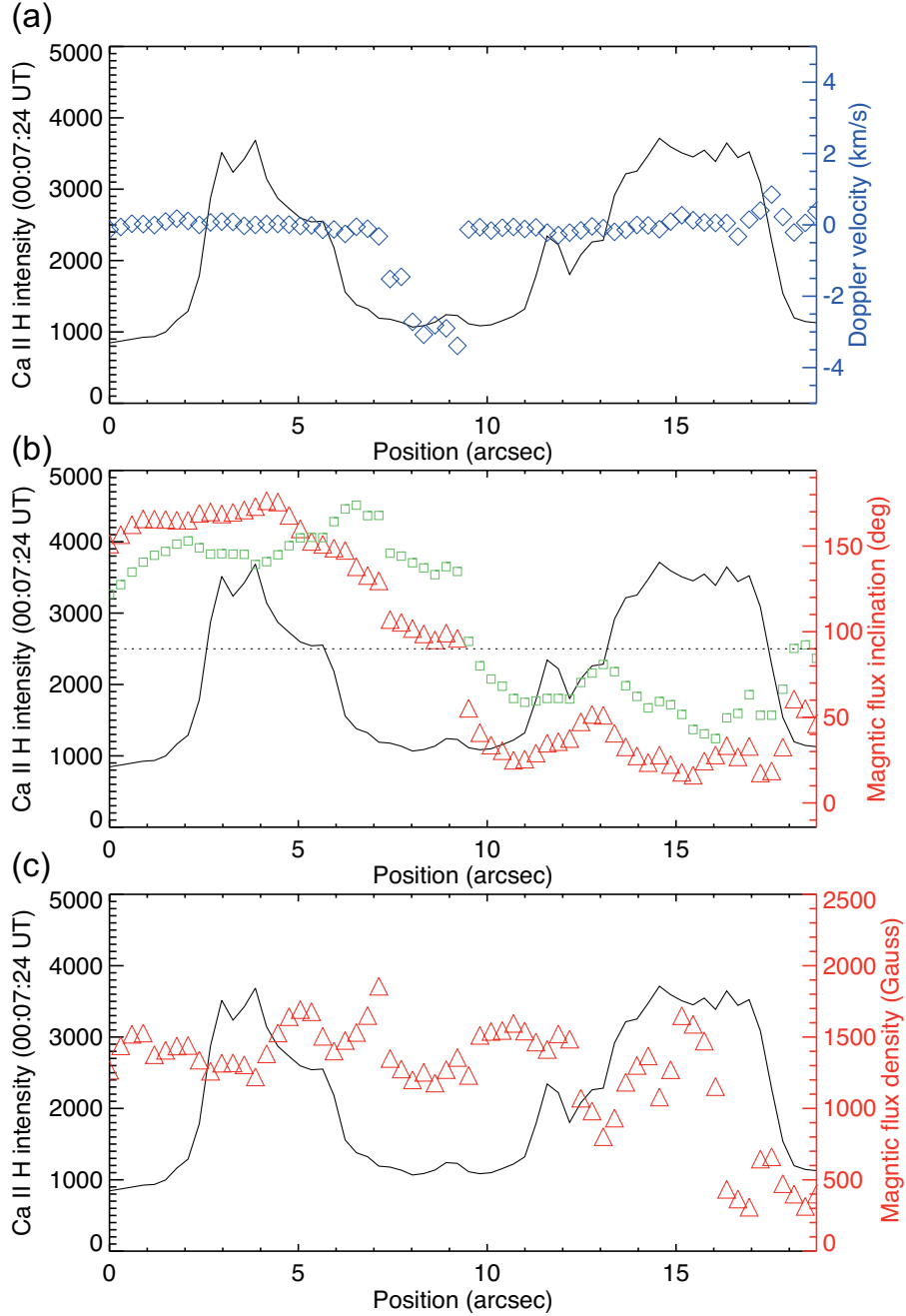} 
 \end{center}
\caption{The physical quantities on 
the dashed line
that passes through the middle 
of the field of view in Figure~\ref{fig: doppler}. a) Doppler velocity (blue 
diamonds) with Ca II H intensity. 
Redshift is positive.
b) Inclination angle of magnetic flux with
Ca II H intensity. Red triangles give the field inclination in the local solar frame,
whereas green squares give the field inclination in the line-of-sight coordinate. 
c) Magnetic flux density (red triangles) with Ca II H intensity. 
The Ca II intensity (solid line in each frame) gives the location of flare ribbons.}
\label{fig: slice}
\end{figure}

Figure~\ref{fig: slice} shows the profiles of the Doppler velocity and
magnetic field quantities on 
the dashed line
that passes through the middle
of the field of view in Figure~\ref{fig: doppler}. Doppler shift signals corresponding 
to 2.5-3.2 km s$^{-1}$ are observed between the flare ribbons.
They are observed to be co-spatial with the horizontally-oriented magnetic field.
As shown by red triangles in Figure~\ref{fig: slice}(b),
the inclination of the magnetic field associated with the Doppler shift
signals  is 95-105\degree~in the local solar frame, where 90\degree~is
horizontal field, i.e.,the field  in parallel to the solar surface. It should be noted that
horizontally-oriented field has an inclination of 132-142\degree~in 
the observer's frame. This inclination is 38-48\degree~from the line-of-sight
direction, so that the uni-directional material flow in the horizontal field can show
Doppler shift signals. Assuming that material flow is aligned to
the horizontal, the actual flow speed may be 4.0-4.3 km s$^{-1}$. 
The magnetic flux density of the horizontal field with the material flow is 1200-1400
Gauss.

\begin{figure}
 \begin{center}
  \includegraphics[width=16cm]{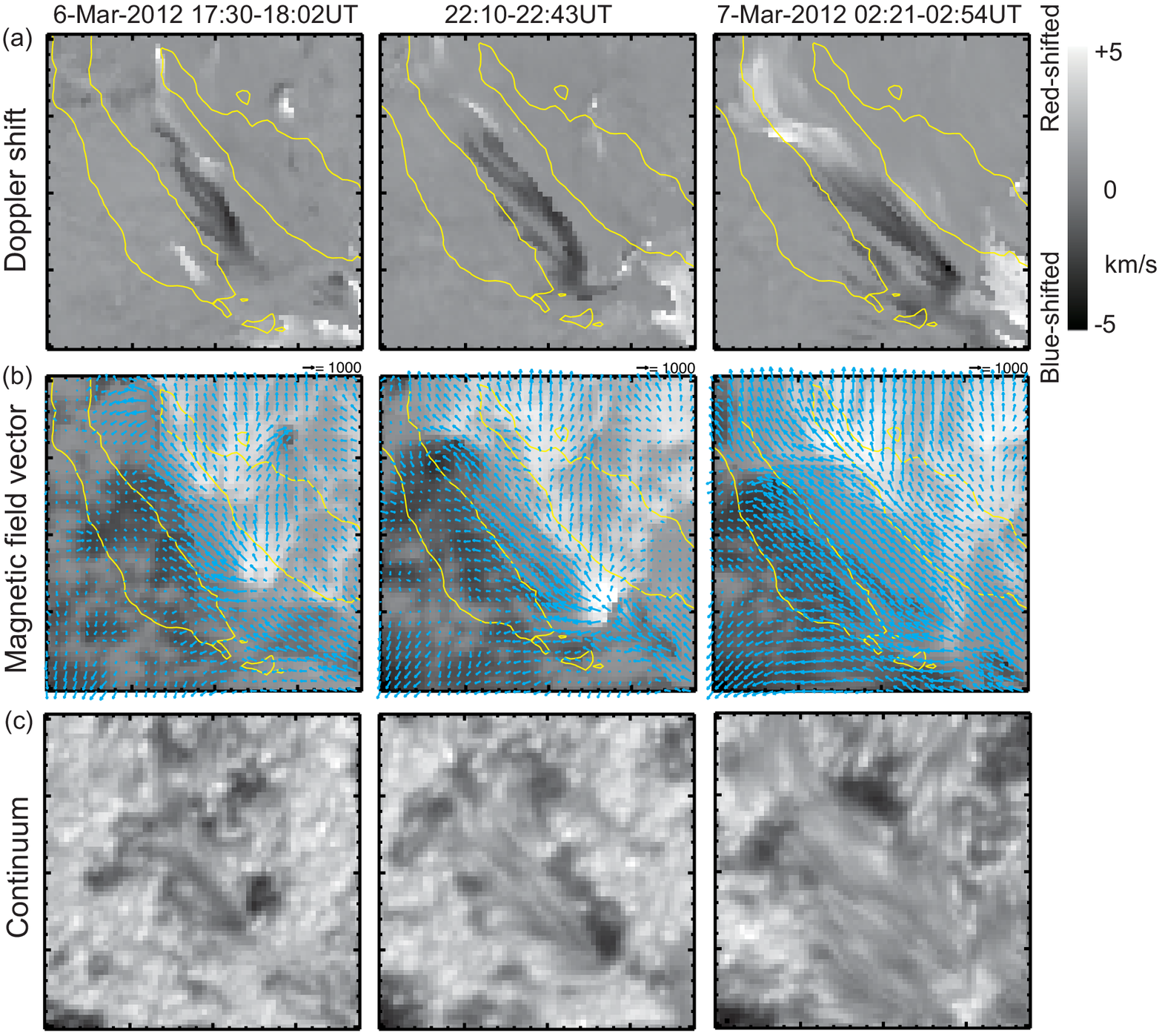} 
 \end{center}
\caption{Temporal evolution of the main energy release site at the photosphere.
(a) Doppler shift signals, (b) vector magnetic field, and (c) continuum images.
The parameters were derived from three SP maps acquired at
17:30-18:20 UT and  22:10 - 22:43 UT on 6 March 2012, and 2:21-2:54 UT on 7 March 2012. 
Yellow contours  show the approximate position of the flare ribbons,
determined from the Ca II H image at 00:07:25 UT, 7 March 2012. 
Redshift is positive in the Doppler signal maps.
On the vector magnetic field maps, the 
aqua-colored arrows show the horizontal component (Bx, By) of 
magnetic field with the gray scale for the vertical component (Bz). 
The field of view, outlined by the square in Figure~\ref{fig: ar}(d),
is about $20 \times 20$ arc sec with 1 arc sec interval between
tick marks .  
North is up and west is to the right.
}
\label{fig: evolution}
\end{figure}

Figure~\ref{fig: evolution} shows  the temporal evolution of the Doppler 
shift signals, vector magnetic field, and photospheric morphology seen in continuum images. 
The three times correspond to about 6 hours and 2 hours before the onset of the X5.4 flare, 
and about 2 hours after the onset of the flare, respectively.  
The high-speed material flow is observed at least 6 hours before the flare onset time. 
It continues to develop even after the onset of the flare.
Simultaneously, the area of the horizontally oriented field gradually increases
with time through the sequence of the three maps. 
A positive polarity magnetic island exists at the SW end of the horizontal field,
whereas a negative polarity  island is observed at the NE end of the horizontal field.

A quantitative measure of these magnetic fields is listed in 
Table~\ref{tab: flux}. 
The inclination ($\gamma$) of the horizontal field 
is on the local solar frame coordinate, in which
90\degree~means the field parallel to the solar surface. 
The error values give the standard deviation of the field inclination distributed 
in the horizontal field region.
The magnetic flux density ($fB$) is the average measured
in the middle area of the horizontal field region with the standard deviation
of the flux density in the measured area.
Since the magnetic flux is horizontally oriented to the solar surface, 
we estimated the magnetic flux ($\phi$) crossing the line-forming height of the Fe I lines
by 
\begin{equation}
 \phi = \int_{S} \vec B \cdot d \vec S = \int_{0}^{w} fB(x) \cos(90-\gamma(x)) dx\ z  .
\end{equation}
For the cross section $S$ of the magnetic flux, we used the measured
width ($w$) of the horizontal flux region and adapted 100 km
as the vertical extent ($z$) of the magnetic flux. 
It is roughly same as the scale height at the photosphere. 
The magnetic flux does not change from 17 UT to 22 UT, but it is increased
by 2.6 times at 2 UT. 
The azimuthal direction of the horizontal field does not change with time and
it is almost in parallel to the polarity inversion line.  
Note that 0\degree~ is the direction toward the west
and 90\degree~ is toward the north. The orientation of 
the flare ribbons (Figure~\ref{fig: evolution}) is aligned almost in parallel
to the direction of the horizontal field.

\begin{figure}
 \begin{center}
  \includegraphics[width=10cm]{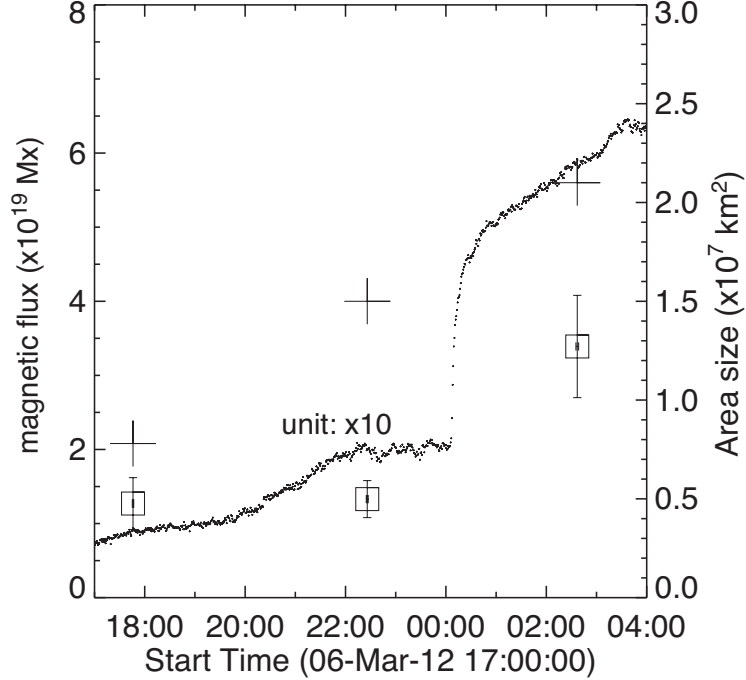} 
 \end{center}
\caption{
The evolution of the region area and magnetic flux of horizontal magnetic
field parallel to the polarity inversion line, with the temporal evolution 
of line-of-sight magnetogram signals for the region of interest, recorded 
in high cadence by HMI. The crosses give the region area and 
the squares are the magnetic flux of the horizontal field. The scale in
the vertical axis is for the magnetic flux of the horizontal field.
The HMI line-of-sight magnetogram signals are the integration of
the magnetic flux density over the region of interest and the vertical
scale for the magnetic flux is $\times 10$ of the unit in the vertical axis.
Since the direction normal to the solar surface
is about 30\degree~ tilted from the line-of-sight direction and 
the enlarging region of
the horizontal field is observed as the region of 
the negative-polarity magnetic flux in the line-of-sight magnetograms.
}
\label{fig: flux}
\end{figure}

The evolution of the magnetic flux and region size is plotted in
Figure~\ref{fig: flux} with the temporal evolution of line-of-sight
magnetogram signals for the region of interest, recorded in high 
cadence by HMI. 
The HMI magnetogram evolution shows a small increase
in the period before the onset of the X5.4 flare and a significant increase
from the flare onset time. 
Since the line-of-sight direction is 
about 30\degree~ tilted from the direction normal to the solar surface
at the region of interest, the enlarging region of
the horizontal field is observed as the region of 
the negative-polarity magnetic flux in the line-of-sight magnetograms.

Table~\ref{tab: flux} also shows the magnetic flux involved in each magnetic flux island
for the three maps.
Since the positive polarity island is distinct relative to its surroundings in 
the magnetic flux density map, the magnetic flux involved in the positive polarity 
island is well-measured. 
It is noted that the measured values of flux for the negative polarity flux island
may contain small amounts of flux from its surroundings because of its
indistinct boundary.
The positive polarity flux 
shows a slight increase
from 17 UT to 22 UT, but after
the X5.4 flare the
flux has reduced to one 
eighth
of the pre-flare flux. 
The separation between the magnetic islands increases slowly 
with time. The speed of separation is 0.13 km s$^{-1}$ in the period from 17 to 22 UT
and 0.04 km s$^{-1}$ in the period from 22 to 2 UT. 
The continuum images reveal penumbra-like structures (orphan penumbra) 
under development,  and these are associated with high-speed material flow.


\begin{table}
  \caption{Quantitative measure of  horizontally oriented field and 
     magnetic islands at the ends of the horizontal field.}\label{tab: flux}
  \begin{center}
    \begin{tabular}{lccc}
      \hline
   & 17:38-17:40 UT & 22:18-22:21 UT &  2:29-2:32 UT \\
      \hline
    Horizontal field region \\
     \hspace{5pt} Region size ($\times 10^{7}$ km$^{2}$) & 0.8 & 1.5 & 2.1 \\
     \hspace{5pt} Flux inclination (\degree) & $89.6\pm 18.5$ & $90.5\pm 20.3$ & 
           $90.8\pm 10.6$ \\  
     \hspace{5pt} Magnetic flux density (Gauss) & $1040\pm250$ & $1050\pm300$ &
            $1500\pm200$ \\
     \hspace{5pt} Magnetic flux ($\times 10^{19}$ Mx) & $1.3\pm0.4$ & $1.3\pm0.3$ &
            $3.4\pm0.7$ \\
      \hspace{5pt} Flux azimuth (\degree) & $138.5\pm 13.1$ & $140.3\pm 10.1$ &
            $141.7\pm 6.7$ \\      
      \hline
    Magnetic islands &  &  &   \\
        \hspace{5pt} Positive polarity's flux ($\times 10^{19}$ Mx)  & $6.8\pm1.0$ &
             $8.2\pm0.5$ & $1.1\pm0.9$ \\
        \hspace{5pt} Negative polarity's flux ($\times 10^{19}$ Mx) & $-14.5\pm1.9$ &
              $-21.8\pm2.5$ & $-10.4\pm1.0$ \\
      \hline
    Separation between the islands (arc sec) & 9.4 & 12.5 & 13.3 \\ 
     \hline
    \end{tabular}
  \end{center}
\end{table}

\section{Discussion}
\label{sec: discussions}

\begin{figure}
 \begin{center}
    \includegraphics[width=8cm]{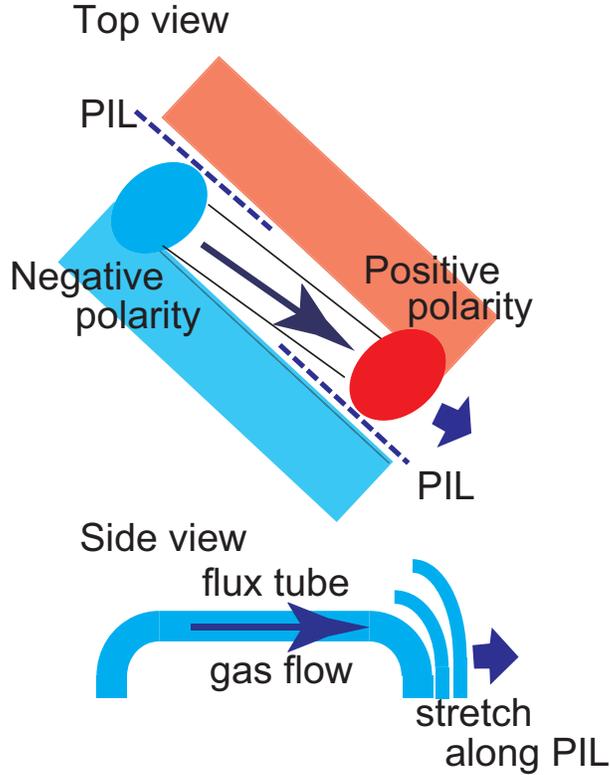} 
 \end{center}
\caption{Summary of observed features around the polarity inversion line (PIL)
 at the photosphere (top view).  In the side view, magnetic field configuration
 with material flow is shown by interpreting the observed features. }
\label{fig: cartoon}
\end{figure}

We observed high-speed Doppler shifts along the polarity inversion line located 
between the flare ribbons of an X5.4 flare occurred on 7 March 2012. 
Observed features around the polarity inversion line in the photosphere 
are summarized in Figure~\ref{fig: cartoon}.
A pair of small magnetic islands exist at the polarity inversion line. The positive polarity
island is located on the positive-polarity side and the negative polarity island is
located on the negative-polarity side.
The orientation of the line defined by the positive and negative polarity islands
is directed nearly along the polarity inversion line. 
Magnetic fields parallel to the solar surface form between the islands.
A high-speed uni-directional material flow 
persists in this horizontal field 
between the islands.
The horizontal field is also associated with the development of 
penumbra-like features in continuum images.
The separation between the magnetic islands increases slowly with time
but the separation speed is slow (0.13 km s$^{-1}$ to 0.04 km s$^{-1}$).
The positive-polarity island located at the end point of the high speed material flow
shows  decrease in magnetic flux -- the largest decrease is observed
in the latter 4-hour interval 
during which
the X5.4 flare occurred.

The slow separation speed
suggests that
the development of horizontally-oriented field between the magnetic islands
is not simply due to the emergence of new magnetic flux from below the photosphere.
The temporal evolution of magnetic flux derived from the HMI line-of-sight
magnetograms shows a small increase in the period before the time of the flare onset, 
but the increase may be artificial because of the flux estimate
only with the line-of-sight component;  
The magnetic flux parallel to the solar surface
is observed as the region of the negative-polarity magnetic flux in
the line-of-sight magnetograms. 
Since the region covered by the horizontal field
is enlarged with time, it results in the apparent increase of 
the magnetic flux.
The two SP measurements before the onset time
clearly show no increase in the magnetic flux of the horizontally oriented
magnetic field at the photospheric level.
A small amount of increase  is seen 
in the magnetic flux of the magnetic islands at the both ends of 
the horizontal field
(
$1.4 \times 10^{19}$ Mx and
$7.3 \times 10^{19}$ Mx for the positive- and negative-polarity, respectively).  
When we assume that  the successive emergence of the horizontal field 
($1.3 \times 10^{19}$ Mx for a vertical thickness of 100 km)
are evolved to the more vertical field at the both ends and added to
the magnetic islands, the emergence up to the height of 
only 110-560 km is expected in the period between the two 
successive SP maps (281 min);
The speed of the emergence is 
0.006-0.033 km/s, which is about one order of magnitude smaller than 
the separation speed of the islands (0.13 km s$^{-1}$)
as well as about two orders of magnitude smaller than the typical speed of flux
emergence at the photospheric level \citep{ots11, shi02}. 

A significant enhancement of the magnetic flux in the horizontal magnetic field
region was detected in the SP data acquired after the onset of the X5.4 
flare. The time series of the HMI line-of-sight magnetograms shows that a rapid 
and significant increase of the negative-polarity signals in the region of interest
has started from the time of the flare onset.  
A similar behavior showing a rapid and irreversible development of
more horizontal orientation of the photospheric magnetic field at 
the flaring magnetic polarity inversion line after flares has been reported
in many observations (e.g., \cite{wan12} and hereinafter); it is interpreted 
that the photospheric magnetic field near the polarity inversion line becomes
more horizontal after eruptions, which could be related to
the newly formed low-lying fields during flares.
It, however, should be noted that the magnitude of the magnetic flux derived from 
the HMI line-of-sight magnetograms contains a large amount of uncertainty,
compared with the magnetic flux from the SP measurements, because of  
the measurements of the only line-of-sight component of the magnetic flux.   

The speed of the separation between the magnetic islands is slow (0.13-0.04 km s$^{-1}$),
but high-speed (about 4 km s$^{-1}$) flows are observed in the horizontal field existing 
between the magnetic islands. 
These high-speed flows are interpreted as one-directional material flows excited 
in the magnetic channels  which are horizontally oriented and connected between
the magnetic islands. 
With the conclusion that the development of the horizontal magnetic field
along the polarity inversion line is not due to a simple emergence of
new magnetic flux from below the photosphere,
the forces caused by emergence are not the primary cause of the one-directional 
material flows.
The flows may be rather driven by the magnetic force given to the horizontal field.
The existence of such flows can be a signature indicating that the magnetic field
near the polarity inversion line is highly stressed,
as shown in some numerical simulations
(e.g., \cite{man01, fan01, mag03, fan12}).

The high-speed flows are observed only in the horizontal field and
they are not observed in the positive-polarity flux concentration 
at the destination of the flows.  
The magnetic field from the flux concentration is extended from the photosphere
to the upper atmosphere because of more vertical orientation, 
whereas the magnetic channels harboring the high-speed
flows may turn back toward the solar interior deep in the atmosphere
at the outer reaches of the horizontal field,
where the flow may stretch the magnetic field and
work as a shear flow along 
the polarity inversion line, rather than a flow converging toward it.
The increasing separation of the magnetic islands enhances the shear in 
magnetic field because the positivity polarity island is located at 
the positive-polarity side and the negative polarity island is located 
at the negative-polarity side of the polarity inversion line. 
This may suggest further development of 
the non-potentiality of the magnetic structure.
However, the increase of shear in magnetic field
may affect a limited portion of the magnetic flux involved in the flare
because the length of the separation between the magnetic islands
is $7,000-10,000$ km, much shorter than the length of the flare ribbons.
High-speed uni-directional material flow along the horizontal field may stretch 
 the magnetic field and cause the positive polarity island to move along 
 the polarity inversion line.
 The kinetic energy of the material flow is estimated to be of order 
$10^{29}$ ergs, which may be sufficient energy to stretch and 
apply the shear force to the magnetic field in the high $\beta$ plasma of
the photosphere. In the gas pressure dominant condition, 
the gas flow applies the additional force to the magnetic field and 
may easily move the magnetic flux.

The observed high-speed material flow also may make a significant contribution
to forming the magnetic structures responsible for the onset of flares. Recently,
\citet{kus12} systematically surveyed  the nonlinear evolution of
magnetic structures toward eruption in terms of 
three-dimensional magnetohydrodynamic simulations.  By injecting
a small-scale bipole field near the polarity inversion line 
of a large-scale force-free field structure,
they identified two typical magnetic structures favorable for the onset of 
eruptive flares: the orientation of the small bipole field is opposite of
the major polarity (opposite-polarity (OP) type) or reversed
to the averaged magnetic shear (reversed-shear (RS) type).
{\em Hinode} observations showed that four major flares have either
OP or RS type configurations \citep{bam13}.
The photospheric magnetic flux distribution of the X5.4 flare may be 
classified as the RS type configuration. The pair of the positive and
negative polarity islands with the uni-directional material flow, however, 
is not the small-scale bipole field
in the \citet{kus12} model because the magnetic shear 
in the large-scale magnetic structure (inferred from the spatial 
distribution of flare ribbons in Figure~\ref{fig: ar}(a)) has almost in the same direction
as the orientation of the pair of the positive and negative polarity islands.
The RS type configuration rather suggests that the combination of the positive 
polarity island at the end point of the uni-directional material flow 
with the negative polarity flux distributed
at the SW of the positive polarity island should be considered as 
the small-scale bipole field.  
The positive polarity island at the end point of the uni-directional material flow
is moved in the SW direction slowly and thus come into close contact with 
the negative polarity flux, which may be preferable for magnetic flux cancellation. 
Moreover, this negative polarity flux is located at the end point of
the significant positive (i.e., red-shifted) Doppler signal seen at the lower right edge 
of Figure~\ref{fig: evolution}. 
This significant positive Doppler signal means that a different uni-directional material
flow exists in the magnetic flux system including the negative polarity flux
and that its flow direction is opposite to the uni-directional material flow 
presented in Section \ref{sec: results}. 
Two material flows streaming in the opposite direction to each other
may squeeze together the positive-polarity island and negative-polarity flux,
which is more preferable for magnetic flux cancellation. 
This suggests an important role of the observed high-speed material flow
in creating the magnetic field configuration favorable for the onset of the X5.4 flare.


\section{Summary}
\label{sec: summary}

High-speed photospheric Doppler shifts are found along the polarity inversion line
located between the flare ribbons of an eruptive major flare on 7
March 2012.  They are observed in a horizontally-oriented magnetic field
formed between positive-polarity and negative-polarity magnetic
islands. Considering that the line-of-sight view for this field nearly parallel to the solar surface, 
the observed Doppler shifts are interpreted as a uni-directional material flow
and the flow is observed to persist from at least 6 hours before 
to at least several hours after the onset of
the flare. 
The temporal evolution of 
the magnetic flux
suggests that the development of the horizontally-oriented
field associated with the material flow is not simply due to the emergence of
new magnetic flux from below the photosphere. The material flow is not a
converging flow toward the polarity inversion line
but it 
can
be recognized as
a flow increasing the shear in magnetic flux. Also, the material flow 
may have
enough kinetic energy to 
move the magnetic flux more easily in the high $\beta$ plasma at the photosphere
and to form 
the magnetic structures favorable for the onset of the eruptive flare. 

The polarity inversion line contains key information for the build-up 
of magnetic energy and the
triggering of solar flares. In addition to the evolving magnetic field configuration, 
material flows around the polarity inversion line are one important 
factor for creating and driving flares. Many more observational examples are
needed to establish a statistical basis for revealing general characteristics of material flows around
the polarity inversion line and their role in the occurrence of flares.
We expect that the {\em Hinode} Solar Optical Telescope and other 
instruments will acquire more good examples of the
photospheric magnetic and velocity fields at the polarity inversion line
of major flares during the maximum phase of the current solar cycle.

\bigskip

We would like to express our thanks to the anonymous referee for 
fruitful  comments.
Hinode is a Japanese mission developed and launched by ISAS/JAXA, with NAOJ
as domestic partner and NASA and STFC (UK) as international partners. It is
operated by these agencies in co-operation with ESA and NSC (Norway). 
We also gratefully acknowledge NASA's Solar Dynamics Observatory and
the HMI science team for providing the magnetograms.
This work was supported by JSPS KAKENHI Grant Number 23540278 and
JSPS Core-to-Core Program 22001.



\begin{thebibliography}{}

\bibitem[Amari et al.(2000)]{ama00}
Amari, T., Luciani, J. F., Mikic, Z., Linker, J.\ 2000, ApJ, 529, L49

\bibitem[Archontis \& Hood(2012)]{arc12}
Archontis, V., Hood, A. W.\ 2010, \aap, 514, A56

\bibitem[Bamba et al.(2013)]{bam13}
Bamba, Y., Kusano, K., Yamamoto, T.T., Okamoto, T. J.\ 2013, ApJ, 778, 48


\bibitem[Canfield et al.(1999)]{can99}
Canfield, R. C., Hudson, H. S., McKenzie, D. E. 1999, Geophys. Res. Lett.,
26, 627

\bibitem[Deng et al.(2006)]{den06}
Deng,N., Xu, Y., Yang, G., Cao, W.,  Liu, C., Rimmele, T.R.,
Wang, H., Denker, C.\ 2006, ApJ, 644, 1278

\bibitem[Fan(2001)]{fan01} 
Fan, Y.\  2001, \apj, 554, L111

\bibitem[Fang et al.(2012)]{fan12} 
Fang, F., Manchester, W., IV, Abbett, W.P., van der Holst, B.\  2012, \apj, 754, 15

\bibitem[Golub et al.(2007)]{gol07} 
Golub, L. et al. 2007, \solphys, 243, 63
  
\bibitem[Ichimoto et al.(2008)]{ich08} 
  Ichimoto, K. et al. 2008, \solphys, 249, 233

\bibitem[Kano et al.(2008)]{kan08} 
  Kano, R. et al. 2008, \solphys, 249, 263

\bibitem[Kosugi et al. (2007)]{kos07} 
  Kosugi, T. et al. 2007, \solphys, 243, 3

\bibitem[Kusano et al. (2012)]{kus12} 
  Kusano, K., Bamba, Y., Yamamoto, T.T., Iida, Y., Toriumi, S., Asai, A.\ 2012, ApJ, 760, 31
  
\bibitem[Lites et al.(1995)]{lit95}
Lites, B. W., Low, B. C.,Mart?Lnez Pillet, V., Seagraves, P., Skumanich, A.,
Frank, Z. A., Shine, R. A., Tsuneta, S.\ 1995, ApJ, 446, 877

\bibitem[Lites et al.(2002)]{lit02}
Lites, B. W., Socas-Navarro, H., Skumanich, A., Shimizu, T.\ 2002, ApJ, 575, 1131

\bibitem[Lites et al.(2013)]{lit13}	
Lites, B. W., Akin, D. L., Card, G., Cruz, T., Duncan, D. W., Edwards, C. G., Elmore, D. F., 
Hoffmann, C., Katsukawa, Y., Katz, N., Kubo, M., Ichimoto, K., Shimizu, T., Shine, R. A., 
Streander, K. V., Suematsu, A., Tarbell, T. D., Title, A. M., Tsuneta, S.\ 2013,
\solphys, 283, 579

\bibitem[Manchester(2001)]{man01}
Manchester, W., IV\  2001, ApJ, 547, 503

\bibitem[Magara \& Longcope(2001)]{mag01}
Magara, T., Longcope, D. W. \  2001, ApJ, 559, L55

\bibitem[Magara \& Longcope(2003)]{mag03}
Magara, T., Longcope, D. W. \  2003, ApJ, 586, 630

\bibitem[Mart\'inez Pillet et al.(1994)]{mar94}
Mart\'inez Pillet, V., Lites, B. W., Skumanich, A., Degenhardt, D. 1994,
ApJ, 425, L113

\bibitem[Meunier \& Kosovichev(2003)]{meu03}
Meunier, N., Kosovichev, A.\ 2003, \aap, 412, 541

\bibitem[Otsuji et al.(2011)]{ots11}
Otsuji, K., Kitai, R., Ichimoto, K., Shibata, K. 2011, \pasj, 63, 1047

\bibitem[Pesnell et al.(2012)]{pes12}
Pesnell, W. D., Thompson, B. J., Chamberlin, P. C. 2012, \solphys, 275, 3

\bibitem[Rust \& Kumar(1996)]{rus96}
Rust, D. M., Kumar, A.\ 1996, ApJ, 464, L199

\bibitem[Sammis, Tang \& Zirin (2000)]{sam00}
Sammis, I., Tang, F., Zirin, H.\ 2000, ApJ, 540, 583

\bibitem[Scherrer et al.(2012)]{sche12}
Scherrer, P.H., Schou, J., Bush, R.I. et al. 2012, \solphys, 275, 207

\bibitem[Schou et al.(2012)]{scho12}
Schou, J., Scherrer, P.H., Bush, R. I. et al. 2012, \solphys, 275, 229

\bibitem[Shimizu et al.(2002)]{shi02}
 Shimizu, T., Shine, R.A., Title, A.M., Tarbell, T.D., Frank, Z. 2002, \apj, 574, 1074

\bibitem[Shimizu et al.(2008)]{shi08} 
  Shimizu, T. et al. 2008, \solphys, 249, 221
    

\bibitem[Suematsu et al.(2008)]{sue08} 
  Suematsu, Y. et al. 2008, \solphys, 249, 197
  
\bibitem[Takizawa et al.(2012)]{tak12}
Takizawa, K., Kitai, R. Zhang, Y.\ 2012, \solphys, 281, 599

\bibitem[Tsuneta et al.(2008)]{tsu08} 
  Tsuneta, S. et al. 2008, Solar Physics, 249, 167
  
\bibitem[Tsurutani et al.(2014)]{tsu14}
Tsurutani, B.T., Echer, E., Shibata, K., Verkhoglyadova, O.P.,
Mannucci, A.J., Gonzalez, W.D., Kozyra, J.U., P\"atzold, M. \ 2014, 
J. Space Weather Clim., 4, A02

\bibitem[van Ballegooijen \& Martens (1989)]{van89}
van Ballegooijen, A. A., Martens, P. C. H.\ 1989, ApJ, 343, 971

\bibitem[Wang et al.(2012)]{wan12}
Wang, S., Liu, C., Liu, R., Deng, N., Liu, Y., Wang, H.\ 2012, ApJ, 745, L17
 
\bibitem[Yang et al.(2004)]{yan04}
Yang, G., Xu, Y., Cao, W., Wang, H., Denker, C., Rimmele, T. R.\ 2004, ApJ,
617, L151


\end{thebibliography}
\end{document}